\documentstyle[preprint,prl,aps]{revtex}
\begin{document}

% comment the following two lines with preprint or default galley style
%\twocolumn[\hsize\textwidth\columnwidth\hsize\csname %
%@twocolumnfalse\endcsname

\draft
\preprint{CTP-2509}
\title{Simulations of Discrete Quantum Systems in Continuous Euclidean 
Time}
\author{B.B. Beard and U.-J. Wiese}
\address{
Center for Theoretical Physics, \\ 
Laboratory for Nuclear Science, and Department of Physics \\
Massachusetts Institute of Technology \\
Cambridge, Massachusetts 02139, U.S.A. \\
}
\date{\today}
\maketitle
\begin{abstract}
Path integrals are usually formulated in discrete Euclidean time using the 
Trotter formula. We propose a new method to study discrete quantum 
systems, in which we work directly in the Euclidean time continuum. The 
method is of general interest and can be applied to studies of quantum spin 
systems, lattice fermions, and in principle also lattice gauge theories. 
Here it is applied to the Heisenberg quantum antiferromagnet using a 
continuous-time version of a loop cluster algorithm. The computational 
advantage of this algorithm is exploited to confirm the predictions of 
chiral perturbation theory in the extreme low temperature regime, down to 
$T = 0.01 J$. A fit of the low-energy parameters of chiral perturbation 
theory gives excellent agreement with previous results and with experiments.  
\end{abstract}
\pacs{75.30.D,75.10.J,02.70.Lq,31.15.K,71.10.F}

% comment the following line with preprint or default galley style
%]

\narrowtext

The path integral for a general quantum system is usually formulated in 
discrete Euclidean time. Farhi and Gutmann pointed out that this is not 
necessary, since the path integral is well-defined in continuous time 
\cite{Farhi92}. For example, for a non-relativistic particle in coordinate 
space, each trajectory in the path integral resembles that of a particle 
undergoing Brownian motion. Due to the continuum of position eigenstates, 
defining the corresponding Wiener measure seems to require discrete time. 
However, one can always work in a countable basis of the Hilbert space: one 
can put the particle in a large box and work in the discrete basis of 
momentum eigenstates, for example. The important point is that then the 
trajectories in the path integral do not resemble Brownian motion. Instead 
they consist of segments for which the system is in a basis state for a 
finite time, with sporadic discrete transitions between basis states. 
Consequently, for discrete quantum systems the path integral can be 
formulated directly in continuous time. This insight can be applied, for 
example, to quantum spin systems, lattice fermions, and lattice gauge 
theories with a compact gauge group.

In the continuous-time formulation a path is characterized by transition 
times and information about which states are connected at the transitions. 
This picture allows the path integral to be sampled numerically without 
having to store information about the individual time slices whose number 
diverges in the time-continuum limit. In the discrete-time approximation
one must always exercise care in extrapolating the results to the 
time-continuum limit. The approach advocated here cuts this Gordian knot by 
operating directly in this limit. This completely eliminates the most 
severe systematic error in these calculations.

As an example we consider quantum spin systems. Conventional approaches to 
handling these systems rely on a discrete-time formalism. Consider, as a 
concrete example, the Hamiltonian for the Heisenberg quantum antiferromagnet
\begin{equation}
H = J \sum_{x,\mu} \vec S_x \cdot \vec S_{x+\hat\mu},
\end{equation}
where $\vec S_x = \frac{1}{2} \vec \sigma_x$ is a spin 1/2 operator 
associated with the site $x$ of a $d$-dimensional hypercubic lattice with 
spacing $a$. The interaction is between nearest neighbors;
$\hat\mu$ is the unit vector in the $\mu$-direction.
For an antiferromagnet the coupling constant $J$ is
positive. Since this Hamiltonian comprises non-commuting terms, the explicit 
evaluation of the associated partition function is problematical. Suzuki 
is credited with showing how the Trotter formula can be applied to segregate 
these non-commuting terms into separate time slices, with the approximation 
becoming exact in the continuum limit \cite{Suzuki76}. For example, the
Hamiltonian of a 1D quantum spin chain can be decomposed into
two terms, $H = H_1 + H_2$, each of which comprises only commuting terms
\begin{equation}
H_1 = J \sum_{x=2m} \vec S_x \cdot \vec S_{x+\hat 1}, \,\,\,
H_2 = J \sum_{x=2m+1} \vec S_x \cdot \vec S_{x+\hat 1}.
\end{equation}
Then for the partition function one writes
\begin{eqnarray}
Z&=&\mbox{Tr} \exp(- \beta H) \nonumber \\
&=&\lim_{N \rightarrow \infty} \mbox{Tr}
[\exp(- \varepsilon \beta H_1) \exp(- \varepsilon \beta H_2)]^N
\end{eqnarray}
where $\beta=1/T$ is the inverse temperature and
$\varepsilon \beta = \beta / N$ is the lattice spacing in Euclidean time.
Inserting complete sets of 
eigenstates $|1\rangle$ and $|-1\rangle$ of $\sigma_x^3$ between the factors
$\exp(- \varepsilon \beta H_i)$ one converts the partition function into
a $(d+1)$-dimensional path integral of Ising-like variables
\begin{equation}
Z = \prod_{x,t} \sum_{s(x,t) = \pm 1} \exp(-S).
\end{equation}
The action is a sum of plaquette couplings
\begin{eqnarray}
S&=&\!\!\!\!\sum_{x=2m,t=2p} \!\!\!\!\!
S[s(x,t),s(x+\hat 1),s(x,t+1),s(x+\hat 1,t+1)]
\nonumber \\
&&\!\!\!\!\!\!\!\!\!\!\!\! + \sum_{x=2m+1,t=2p+1} \!\!\!\!\!\!\!\!\!\!\!\!
S[s(x,t),s(x+\hat 1),s(x,t+1),s(x+\hat 1,t+1)], \nonumber \\ \
\end{eqnarray}
and the plaquette Boltzmann factors are given by a $4 \times 4$ transfer
matrix
\begin{equation}
\exp(-S[s_1,s_2,s_3,s_4]) = \langle s_1 s_2 |
\exp(- \varepsilon \beta J \vec S_x \!\!\cdot \!\!\vec S_{x+\hat\mu}) |
s_3 s_4 \rangle.
\end{equation}
For the spin-$1/2$ Heisenberg Hamiltonian, only six of the sixteen 
elements of the transfer matrix are non-zero, namely those that leave the 
$3$-component of spin unchanged. 

The 2D Heisenberg model is of interest because it describes the undoped 
precursor insulators of high temperature superconductors, and it has been 
studied with various Monte Carlo techniques \cite{Makivic91}, including a 
very efficient loop cluster algorithm \cite{Evertz93,Wiese94}. The cluster 
algorithm has led to a very precise determination of the low energy 
parameters of the Heisenberg model \cite{Wiese94}, consistent with the 
chiral perturbation theory treatment of Hasenfratz and Niedermayer 
\cite{Hasenfratz93}. The loop cluster algorithm has a well-defined 
continuous-time limit, as discussed below. The computational advantage of 
operating directly in the time continuum allows us to verify the 
predictions of chiral perturbation theory at temperatures that are 
inaccessible with the conventional discrete-time algorithm. The results for 
the low energy parameters of the theory are consistent with previous 
calculations \cite{Wiese94} as well as with experimental data on precursor 
insulators of high temperature superconductors \cite{Shirane87,Greven95}.  

The loop cluster algorithm in discrete time is implemented on a space-time 
lattice with $(L/a)^2$ spatial points and $4N$ time slices, with a $\pm 1/2$ 
spin state defined at each lattice point. Both in space and in Euclidean
time periodic boundary conditions are used. Decomposing the Hamiltonian 
with the Trotter formula allows us to define an action for each lattice 
configuration. Loops are built up according to a set of rules dictating the 
loop ``flow'' through each plaquette. After a loop is built, every spin on the 
loop is flipped. The loop-building rules are calibrated to satisfy the 
condition of detailed balance, ensuring that the path configuration space is 
sampled in proportion to the statistical weight of each configuration.

The discrete-time rules, together with their continuum analogs, are 
summarized in Table \ref{flowrules}. Time is in the vertical direction for 
the plaquettes displayed here. The essence of the loop cluster algorithm is 
that we assign a ``flow'' to each legal plaquette, that is, to each 
plaquette with a non-zero entry in the transfer matrix and hence with 
finite action. In the case of the ``optional decay'' plaquettes the flow 
can proceed in either the space or the time direction; we choose the 
probability for proceeding in the space direction so that the entire 
algorithm satisfies detailed balance. Starting at a random site in 
space-time, we connect flows until the loop closes, and then all spins on 
the loop are flipped. The cluster-building rules ensure that the cluster 
paths always close. 

The continuum limit is determined by first observing that jumps from 
one spin site to a neighbor (the ``optional'' plaquettes in the table) are 
sporadic and of measure zero, so that most plaquettes are continuations in 
time. The solid lines in Table \ref{flowrules}
represent continuously spin up states, and 
the dotted lines are continuously spin down states. The forced transition 
plaquettes become ``bonds'' in the continuum limit, and the flow is required 
to travel along the bond and reverse direction in the continuum paradigm. 
The optional decay plaquettes become Poisson-like decays between unlike 
spins; the probability of a jump in the discrete case becomes a probability 
per unit time in the continuum case. The duration of a path segment on a 
given spin site is thus distributed like the lifetime of a radioactive 
nucleus: the more decay channels available, the more likely the path will 
terminate in a given amount of time.
The essential difference between the discrete-time and continuous-time 
implementation is that the former requires us to store spin state 
information for each of the points of the space-time lattice, 
while the latter requires us only to store the transition times for each 
spin site (plus an extra bit to record the state at $t = 0$). In the code 
developed here, all the transition times were maintained in a large array. A 
cyclic double-linked-list storage technique was implemented to ensure that 
the overhead associated with inserting and deleting time values was 
minimized.

A significant advantage of the loop cluster algorithm is that it allows for 
the implementation of improved estimators. We found that the improved 
estimators for susceptibility, staggered susceptibility, and internal 
energy density all have easily determined continuum limits. 

The continuous-time algorithm with improved estimators was implemented in a 
FORTRAN code \cite{mynote1}. Instead of the node-by-node method of the 
discrete-time algorithm, the continuous-time algorithm is based on the 
exponential distribution of decay times. Starting at a randomly selected 
space-time location in the lattice, the neighborhood of the current 
endpoint of the path is surveyed. The first job of the survey is to 
determine the ``horizon'', which is the time of the next forced transition. 
The other purpose of the survey is to determine the number and duration of 
decay channels --- i.e. sites with opposite spin unvisited by the current 
path. The time of the nearest change in the neighborhood is called the
``waypoint.'' A pseudo-random number is generated, and a decay-point is 
computed according to the exponential distribution corresponding to the 
number of decay channels. The current time marker is moved to the closer of 
the waypoint or the decay-point. If the marker reaches the waypoint, then 
the waypoint and decay-point determination is repeated. If the marker 
reaches either the horizon or a decay-point, then the path moves to the 
appropriate neighbor, selected randomly from several channels if necessary. 
The process continues until the path closes. 

In general terms, the computer time required to run a simulation for a 
given volume and temperature is roughly comparable for the discrete code 
(at a usable time-granularity) and the continuum code, but the continuum 
code requires substantially less storage space. The real advantage of the 
continuum code is that it obviates the need to conduct the several runs of 
successively finer time-granularity needed for extrapolation to the 
continuum limit, thus obviating a costly dimension in the simulation 
procedure.

In the ground state of the 2D Heisenberg antiferromagnet the staggered 
magnetization develops an expectation value, and hence the $O(3)$ spin 
rotational symmetry gets spontaneously broken down to $O(2)$. The low 
energy excitations of the systems are spinwaves (so-called magnons) which 
are the Goldstone bosons of the spontaneously broken $O(3)$ symmetry. Using 
symmetry arguments, chiral perturbation theory makes very strong 
predictions for magnon dynamics at low temperatures, with three low energy 
parameters as unknown constants. These parameters are the staggered 
magnetization ${\cal M}_s$, the spin wave velocity $\hbar c$, and the spin 
stiffness $\rho_s$. The predictions of chiral perturbation theory at 
extremely low temperatures were checked using a range of square volumes 
with side length $L/a = 6,8 \ldots 20$ and a range of inverse temperatures 
$\beta J = 1, 2, 5, 10, 20, 30, 40, 50, 80, 100$. Note that the very small 
temperatures $T \approx 0.01 J$ are inaccessible to the discrete-time code, 
largely due to storage limitations. In every case 100,000 configurations 
were generated after a thermalization period of 1000 configurations. 

Several of the key predictions of chiral perturbation theory were directly 
verified with this code. In particular, the 
energy spectrum is that of an $O(3)$ rotor, with energy levels characterized 
by an integer spin $j$, having degeneracy $2 j  + 1$  and energy levels 
distributed as $j ( j + 1 )$. Finite volume effects are computed as 
expansions in $\hbar c/\rho_s L$, in units where the lattice spacing $a = 1$. 
The energy spectrum is computed in reference \cite{Hasenfratz93} to be 
\begin{eqnarray}
&&E_j - E_0 = \nonumber \\
&&j (j+1) \frac{(\hbar c)^2}{2 \rho_s L^2} 
\left[ 1 - \frac{\hbar c}{\rho_s L} \frac{3.900265}{4 \pi} + 
O(L^{-2}) \right].
\end{eqnarray}
Including only the leading term in the partition function, the uniform 
susceptibility approaches the form 
\begin{equation}
\label{chiu_eqn}
\chi_u \mathrel{\mathop{\kern0pt\longrightarrow}\limits_{T \to 0}} 
\frac{6}{L^2 T} \exp \left( - \frac{(\hbar c)^2}{\rho_s L^2 T} \right),
\end{equation}
and the staggered susceptibility goes to the temperature-independent form
\begin{equation}
\label{chis_eqn}
\chi_s \mathrel{\mathop{\kern0pt\longrightarrow}\limits_{T \to 0}}
\frac{2 {\cal M}_s^2 \rho_s}{(\hbar c)^2} L^4 
\left[1 + 3 \frac{\hbar c}{\rho_s L}\frac{3.900265}{4\pi}
+ O(L^{-2}) \right].
\end{equation}
These functional forms in the large $\beta$ limit, including both volume and 
temperature dependence, were verified for uniform and staggered 
susceptibility. In addition, the staggered susceptibility is shown to 
plateau at large $\beta$ as predicted by chiral perturbation theory. 

An independent fit for the chiral perturbation theory parameters ${\cal M}_s$,
$\hbar c$, and $\rho_s$ gives
excellent agreement with the results of \cite{Wiese94}, as shown in 
Table \ref{parameters}.
This fit required that the partition function of the $O(3)$ rotor be 
included in its entirety, instead of including just the leading term, 
as the limiting 
forms in equations (\ref{chiu_eqn}) and (\ref{chis_eqn}) employ. In 
addition, only inverse temperatures $\beta J \ge 10$ were used in the fit, 
since the rotor approximation is valid only for very small temperatures. In 
order to reproduce the accuracy of the fit in reference \cite{Wiese94}, it was 
necessary to find fitted values for the coefficients of the quadratic terms 
in the expressions for energy and staggered susceptibility, that is, the 
coefficients of the terms $(\hbar c/\rho_s L)^2$ in the expressions above. 
Note that the agreement between the current work and reference 
\cite{Wiese94} is particularly remarkable because they are derived for 
different volume-temperature regimes. Reference \cite{Wiese94} was concerned 
with the ``cubical'' regime $T L/\hbar c \cong 1$, while the
current study focuses on the ``cylindrical'' regime $T L /\hbar c << 1$. 
Quadratic coefficients for the cubical regime are not included in 
Table \ref{parameters} 
because they are not comparable to those in the cylindrical regime. 
The five parameter values in Table \ref{parameters} resulted in a 
goodness-of-fit $\chi^2/\mbox{d.o.f.}=1.5$.
Figures \ref{chis}(a) and \ref{chis}(b) show the fit for 
uniform and staggered susceptibility, 
respectively. Solid lines represent the fitted chiral perturbation theory 
result. Circles and error bars are displayed at each simulation point. The 
fit is quite good for $\beta J \ge 10$. 

In conclusion, we find that the continuous-time formulation discussed here 
provides a superior method for evaluating path integrals that arise 
in the study of discrete quantum systems. Using this method allows us to 
validate the chiral perturbation theory results for the Heisenberg 
antiferromagnet in the extreme low temperature regime. In discrete
Euclidean time a loop cluster algorithm has also been constructed for 
lattice fermions \cite{Wiese93}. Our continuous time method can be applied
to this problem as well. In fact, in principle it works for any discrete
quantum system, even if a cluster algorithm may not be available. For
example, for a
nonrelativistic particle living in a finite volume one can formulate the
path integral using the discrete momentum states. Similarly, for
a lattice gauge theory with a compact gauge group one can work in the 
discrete basis of representations. In both cases this yields a well-defined 
path integral directly in the Euclidean time continuum.

We would like to thank P. Grassberger for asking a question that led to 
this project. We also thank E. Farhi and S. Gutmann for helpful 
discussions. This material is based upon work supported under a National 
Science Foundation Graduate Fellowship (BBB) and an Alfred P. Sloan 
Fellowship (UJW). This work has also been supported in part by funds 
provided by the U.S. Department of Energy (D.O.E.) under cooperative 
research agreement DE-FC02-94ER40818.

\newpage

\begin{figure}
\caption{(a) Uniform susceptibility
and (b) staggered susceptibility
versus inverse temperature
for various volumes. Solid lines are predictions from chiral 
perturbation theory with fitted parameters; circles with error bars
are lattice simulations with continuous time. The fit is very good
for $\beta J \geq 10$.}
\label{chis}
\end{figure}

\newpage

\begin{table}
\caption{Summary of Plaquette Flow Rules.
$+1/2$ and solid lines indicate spin-up sites;
$-1/2$ and dotted lines are spin-down. The time direction is vertical.
Flow rules for inverse plaquettes are analogous. \label{flowrules}}
\begin{tabular}{p{3.5cm} cc}
% Column Labels
& Discrete Time & Continuous Time\\
\tableline

% Line 1 of table "flowrules"
   \begin{tabular}{l}
   Forced continuation \end{tabular} &
% discrete time forced continuation
   \begin{tabular}{c}
      \begin{picture}(60,70)(-30,-35)
         \put(20,-20){\makebox(0,0){$+\frac{1}{2}$}}
         \put(-20,-20){\makebox(0,0){$+\frac{1}{2}$}}
         \put(20,20){\makebox(0,0){$+\frac{1}{2}$}}
         \put(-20,20){\makebox(0,0){$+\frac{1}{2}$}}
         \put(20,-10){\vector(0,1){20}}
         \put(-20,-10){\vector(0,1){20}}
      \end{picture} \end{tabular} &
   \begin{tabular}{r}
% continuum time forced continuation
      \begin{picture}(60,60)(-30,-30)
      \thicklines
         \put(20,-20){\line(0,1){40}}
         \put(-20,-20){\line(0,1){40}}
      \thinlines
         \put(15,-15){\vector(0,1) {30}}
         \put(-15,-15){\vector(0,1){30}}
      \end{picture} \end{tabular} \\

% Line 2 of table "flowrules"
   \begin{tabular}{l}
   Forced transition \end{tabular} &
% discrete time forced transition
   \begin{tabular}{c}
      \begin{picture}(60,70)(-30,-35)
         \put(20,-20){\makebox(0,0){$+\frac{1}{2}$}}
         \put(-20,-20){\makebox(0,0){$-\frac{1}{2}$}}
         \put(20,20){\makebox(0,0){$-\frac{1}{2}$}}
         \put(-20,20){\makebox(0,0){$+\frac{1}{2}$}}
         \put(10,20){\vector(-1,0){20}}
         \put(10,-20){\vector(-1,0){20}}
      \end{picture} \end{tabular} &
   \begin{tabular}{r}
% continuum time forced transition
      \begin{picture}(60,60)(-30,-30)
      \thicklines
         \put(20,-20){\line(0,1){20}}
         \put(20,0){\line(-1,0){40}}
         \put(-20,0){\line(0,1){20}}
         \multiput(20,5)(0,5){4}{\circle*{2}}
         \multiput(-20,-5)(0,-5){4}{\circle*{2}}
      \thinlines
         \put(15,15){\vector(0,-1){10}}
         \put(15,-15){\vector(0,1){10}}
         \put(-15,5){\vector(0,1){10}}
         \put(-15,-5){\vector(0,-1){10}}
         \put(15,5){\vector(-1,0){30}}
         \put(15,-5){\vector(-1,0){30}}
      \end{picture} \end{tabular} \\

% Line 3 of table "flowrules"
   \begin{tabular}{p{3.3cm}}
   \begin{flushleft}
   Optional decay,
   \mbox{$ p = 2 / ( 1 + \mbox{exp} ( \epsilon \beta J))$}
   and \mbox{$ \lambda = \beta J / 2 $}
   \end{flushleft} \end{tabular} &
% discrete time optional decay
   \begin{tabular}{c}
      \begin{picture}(60,65)(-30,-30)
         \put(20,-20){\makebox(0,0){$-\frac{1}{2}$}}
         \put(-20,-20){\makebox(0,0){$+\frac{1}{2}$}}
         \put(20,20){\makebox(0,0){$-\frac{1}{2}$}}
         \put(-20,20){\makebox(0,0){$+\frac{1}{2}$}}
         \put(20,10){\vector(0,-1){20}}
         \put(-20,-10){\vector(0,1){20}}
         \put(10,20){\vector(-1,0){20}}
         \put(-10,-20){\vector(1,0){20}}
         \put(25,0){\makebox(0,0){$\scriptstyle p$}}
         \put(0,25){\makebox(0,0){$\scriptstyle 1-p$}}
      \end{picture} \end{tabular} &
   \begin{tabular}{r}
% continuum time optional decay
      \begin{picture}(60,60)(-30,-30)
      \thicklines
         \multiput(-20,0)(5,0){8}{\line(1,0){3}}
         \put(-20,-20){\line(0,1){40}}
         \multiput(20,-20)(0,5){9}{\circle*{2}}
      \thinlines
         \put(15,15){\vector(0,-1){10}}
         \put(15,-5){\vector(0,-1){10}}
         \put(-15,5){\vector(0,1){10}}
         \put(-15,-15){\vector(0,1){10}}
         \put(15,5){\vector(-1,0){30}}
         \put(-15,-5){\vector(1,0){30}}
         \put(-2,10){$\lambda$}
      \end{picture} \end{tabular} \\

\end{tabular}
\end{table}

\begin{table}
\caption{Comparison of Fitted Parameters \label{parameters}}
\begin{tabular}{l r@{}c@{}l r@{}c@{}l@{}}
% & Ref \cite{Wiese94} & Current \\
&
\multicolumn{3}{c}{Ref \cite{Wiese94}} &
\multicolumn{3}{c}{Current} \\
\tableline
Spin stiffness $\rho_s$ & 0&.&186(4) & 0&.&185(2) \\
Spin wave velocity $\hbar c$ & 1&.&68(1) & 1&.&68(1) \\
Staggered Magnetization ${\cal M}_s$ & 0&.&3074(4) & 0&.&3083(2) \\
Quadratic coeff.: Energy & & &--- & 0&.&068(1) \\
Quadratic coeff.: Stag. Susc. & & &--- & 0&.&338(7) \\
\end{tabular}
\end{table}

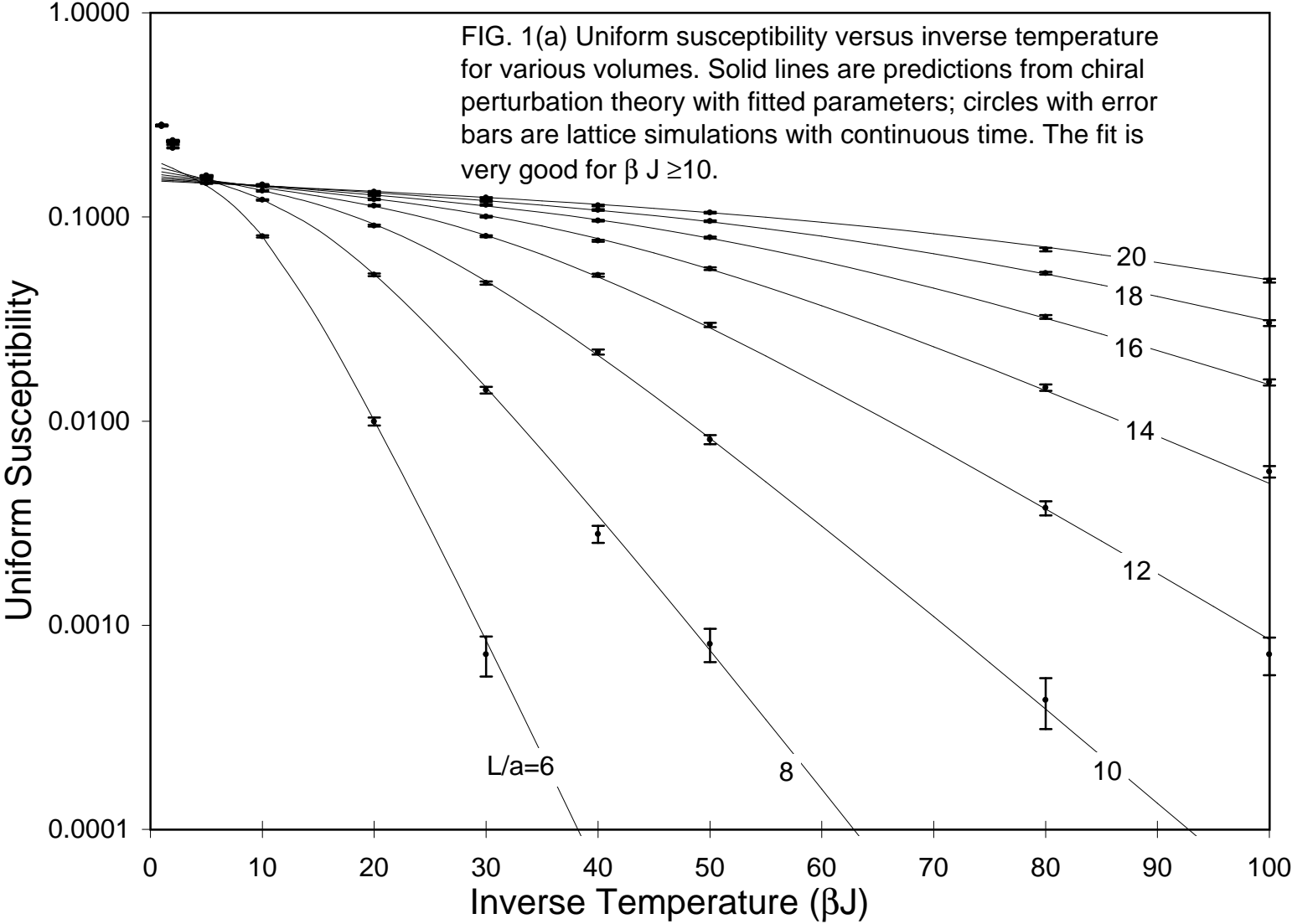
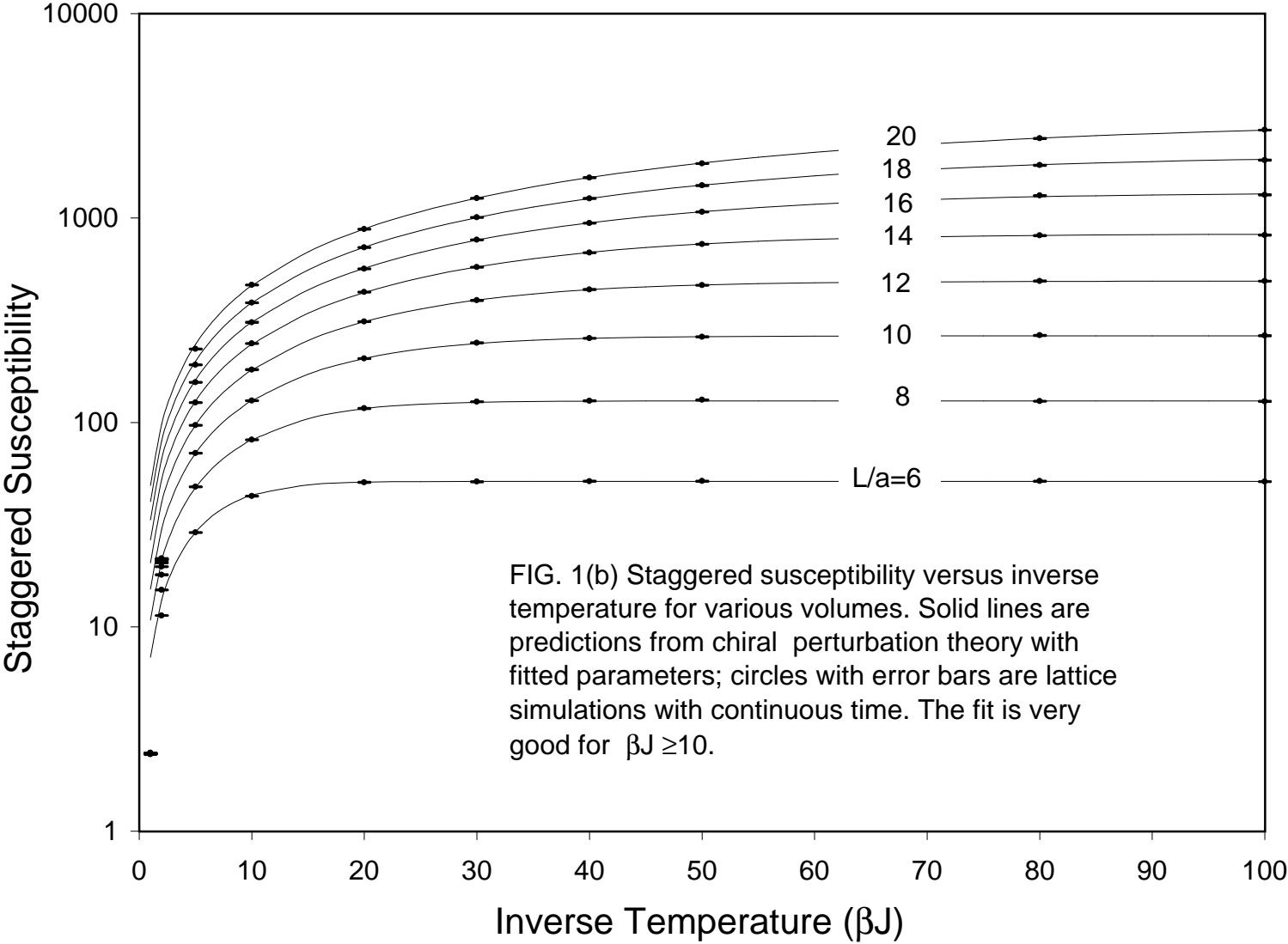
\end{document}